\documentclass[apl, twocolumn,tightenlines]{revtex4-1}

\usepackage{amsthm}
\usepackage{amsmath}
\usepackage{bbm, dsfont}
\usepackage{graphicx}
\usepackage[usenames,dvipsnames]{color}
\usepackage[colorlinks=true,citecolor=magenta,urlcolor=blue]{hyperref}
\usepackage[table]{xcolor}
\usepackage{colortbl}
\usepackage{color}
\usepackage{multirow}
\usepackage{hhline}
\usepackage{tabu}
\usepackage{epstopdf}
\usepackage{booktabs}
\usepackage[normalem]{ulem}
\usepackage{siunitx}
\usepackage{enumitem}
\usepackage{braket}
\usepackage{subcaption}

\usepackage[textwidth=3cm, textsize=footnotesize]{todonotes}

\newcolumntype{C}[1]{ >{ \centering\arraybackslash}p{#1}}

\newcommand{\X}{$\mathsf{X}$}
\newcommand{\Z}{$\mathsf{Z}$}

\newcommand{\nz}{n_\mathsf{Z}}
\newcommand{\nzmu}[1]{n_{\mathsf{Z},\mu_#1}}
\newcommand{\nzk}{n_{\mathsf{Z},k}}

\newcommand{\mz}{m_\mathsf{Z}}
\newcommand{\mzmu}[1]{m_{\mathsf{Z},\mu_#1}}
\newcommand{\mzk}{m_{\mathsf{Z},k}}

\newcommand{\mxmu}[1]{m_{\mathsf{X},\mu_#1}}

\newcommand{\sz}[1]{s_{\mathsf{Z}, #1}}
\newcommand{\sx}[1]{s_{\mathsf{X}, #1}}

\newcommand{\vz}[1]{v_{\mathsf{Z}, #1}}
\newcommand{\vx}[1]{v_{\mathsf{X}, #1}}

\newcommand{\phiz}{\phi_{\mathsf{Z}}}

\definecolor{gray4}{gray}{0.8}
\definecolor{gray2}{gray}{0.4}

\newcommand{\tab}[1]{\begin{tabular}{c} #1 \end{tabular}}
\newcommand{\tabval}[2]{\begin{tabular}{c} #1 \\ {\color{gray2} #2} \end{tabular}}

\begin{document}
\title{Finite-key analysis for the 1-decoy state QKD protocol}

\author{Davide Rusca}\email{davide.rusca@unige.ch}
\affiliation{Group of Applied Physics, University of Geneva, Chemin de Pinchat 22, CH-1211 Geneva 4, Switzerland}
\author{Alberto Boaron}
\affiliation{Group of Applied Physics, University of Geneva, Chemin de Pinchat 22, CH-1211 Geneva 4, Switzerland}
\author{Fadri Gr\"unenfelder}
\affiliation{Group of Applied Physics, University of Geneva, Chemin de Pinchat 22, CH-1211 Geneva 4, Switzerland}
\author{Anthony Martin}
\affiliation{Group of Applied Physics, University of Geneva, Chemin de Pinchat 22, CH-1211 Geneva 4, Switzerland}
\author{Hugo Zbinden}
\affiliation{Group of Applied Physics, University of Geneva, Chemin de Pinchat 22, CH-1211 Geneva 4, Switzerland}


\begin{abstract}
It has been shown that in the asymptotic case of infinite-key length, the 2-decoy state QKD protocol outperforms the 1-decoy state protocol. Here, we present a finite-key analysis of the 1-decoy method. Interestingly, we find that for practical block sizes of up to $10^8$ bits, the 1-decoy protocol achieves for almost all experimental settings higher secret key rates than the 2-decoy protocol. Since using only one decoy is also easier to implement, we conclude that it is the best choice for QKD, in most common practical scenarios.
\end{abstract}

\maketitle

 
Quantum Key Distribution has been originally designed to work with true single-photons \cite{Bennett1984}. However, more than 30 years later, suitable deterministic single-photon sources are still not available.
Therefore in most experimental setups, convenient weak coherent laser pulses are used \cite{Gisin2002,scarani2009}.
Weak coherent pulses are vulnerable to the so called photon number splitting (PNS) attack exploiting multi-photon pulses~\cite{huttner1995,brassard2000}.
This attack can be mitigated using small average photon numbers $\mu$, or particular protocols which are more resistant by design~ \cite{Scarani2004,Inoue2002,Stucki2005}. 
However, arguably the most efficient counter-measure is the so-called decoy-method~\cite{Hwang2003,Lo2004}. 
In this method Alice chooses randomly the average photon number among different levels $\mu_i$ and analyses statistically the probabilities of detection at Bob’s in order to detect a possible PNS attack.

The decoy state protocol was proposed by Hwang~\cite{Hwang2003} and the first complete security proof of the decoy-method was given in 2005 by Lo et al~\cite{Lo2004} for an infinite amount of intensities.Wang~\cite{wang2005} showed, instead, that it was possible to employ the decoy method with only three intensities, i.e. two decoys and one signal state. Later Ma et al.~\cite{Ma2005} demonstrated that in the optimal configuration, one of the two decoys must be set close to the vacuum state (vacuum + weak decoy state protocol).
In the same work, a simpler method with only two intensities was presented as well, i.e. a signal and a decoy states. 
Its security was proved, but the achieved SKR was slightly below the 2-decoy protocol. 
However the analysis did not take into account the statistical correction due to a finite-key length. 
This was first done by Hayashi et al.~\cite{hayashi2014} and then by Lim et al.~\cite{Lim2014}, using a simpler approach, but still only for the 2-decoy configuration.

In this paper, we compare the performance of 1-decoy and 2-decoy levels approaches, following the method used by Lim et al. in 2014. Taking into account finite size effects, we show that, interestingly, for most experimental settings the use of only 1-decoy level is advantageous. 


The previous finite-key analysis of the 2-decoy method, bounded the secret key length of the protocol to the quantity ~\cite{Lim2014}:

\begin{align}\label{eq_skr}
l \leq &  \sz{0}^l + \sz{1}^l(1-h(\phiz^u)) - \lambda_\text{EC} \nonumber\\
& - a\log_2(b/\epsilon_\text{sec}) - \log_2(2/\epsilon_\text{cor}),
\end{align}

\noindent where $\sz{0}^l$ is the lower bound on the vacuum events ($\sz{0}$); those events where Bob had a detection and the pulse sent by Alice contained no photons, $\sz{1}^l$ is the lower bound on the single-photon events ($\sz{1}$), defined by the number of detections at Bob side when the pulse sent by Alice contained only one photon, $\phiz^u$ is the upper bound on the phase error rate ($\phiz$), $\lambda_\text{EC}$ is the number of disclosed bits in the error correction stage, $\epsilon_\text{sec}$ and $\epsilon_\text{cor}$ are the secrecy and correctness parameters and $a$ and $b$ depend on the specific security analysis taken into account ($a = 6$ and $b = 21$ for the 2-decoy approach and $a = 6$ and $b = 19$ for the 1-decoy protocol, see supplementary material for details).

The main contribution to the secret key is given by the single-photon events, estimated by the following formula:
\begin{multline}\label{eq} 
\sz{1} \geq \sz{1}^l := \frac{\tau_1\mu_1}{\mu_1(\mu_2-\mu_3)-\mu^2_2+\mu^2_3}\left(\nzmu{2}^- \right. \\
\left. - \nzmu{3}^+ + \frac{(\mu^2_2-\mu^2_3)}{\mu^2_1}\left(\frac{\sz{0}}{\tau_0} - \nzmu{1}^+\right)\right),
\end{multline}
where $\tau_n$ is the total probability to send an $n$-photon state and $\nzk^\pm$ is the finite-key correction, obtained by using the Hoeffding's inequality~\cite{Hoeffding1963}, of the number of detections in the \Z{} basis due to the state of intensity $k \in \left\lbrace \mu_1, \mu_2, \mu_3\right\rbrace $:
\begin{align}
\nzk^\pm &:=   \frac{e^{k}}{p_k}\left(\nzk \pm \sqrt{\frac{\nz}{2}\log\frac{1}{\varepsilon_1}}\right).
\end{align}

\begin{figure*}
\centering
\begin{minipage}{.5\textwidth}
\includegraphics[width=1\linewidth]{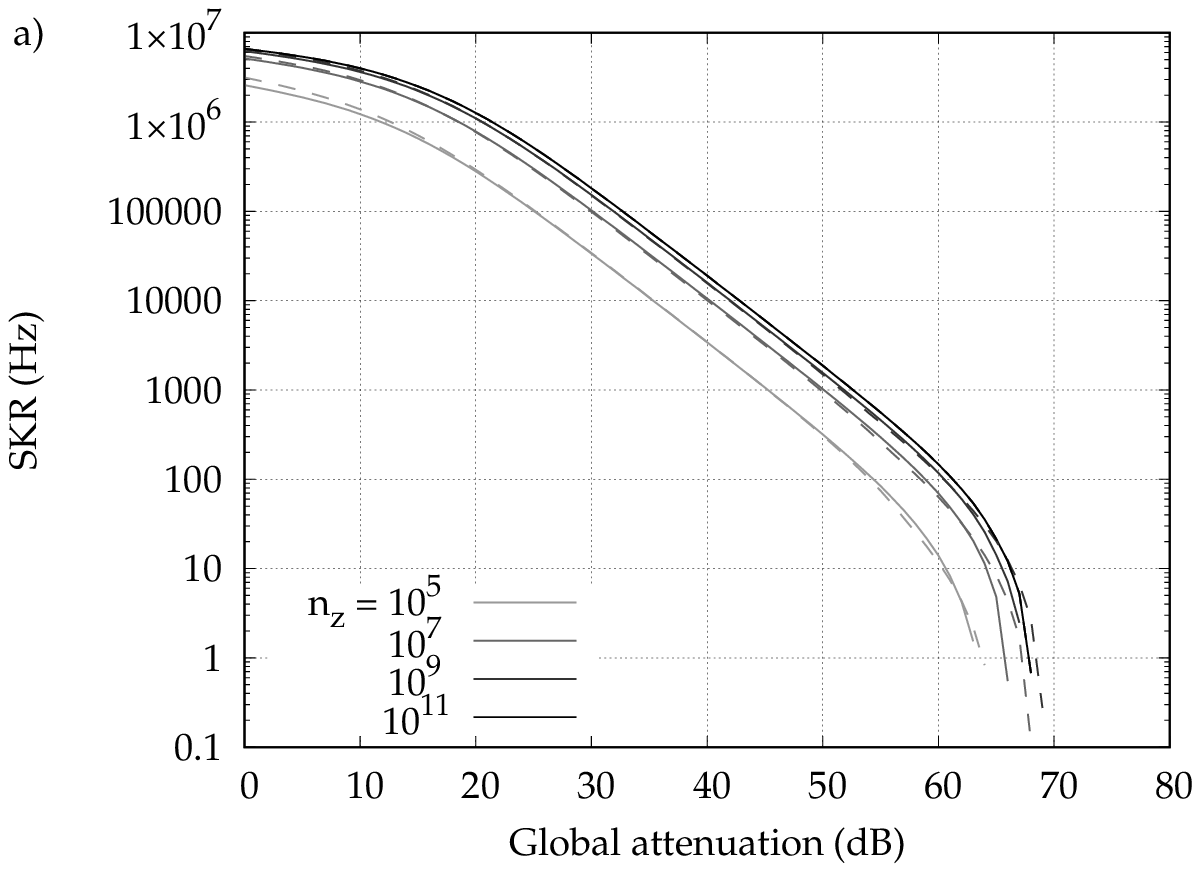}
\end{minipage}%
\begin{minipage}{.5\textwidth}
  \centering
  \includegraphics[width=1\linewidth]{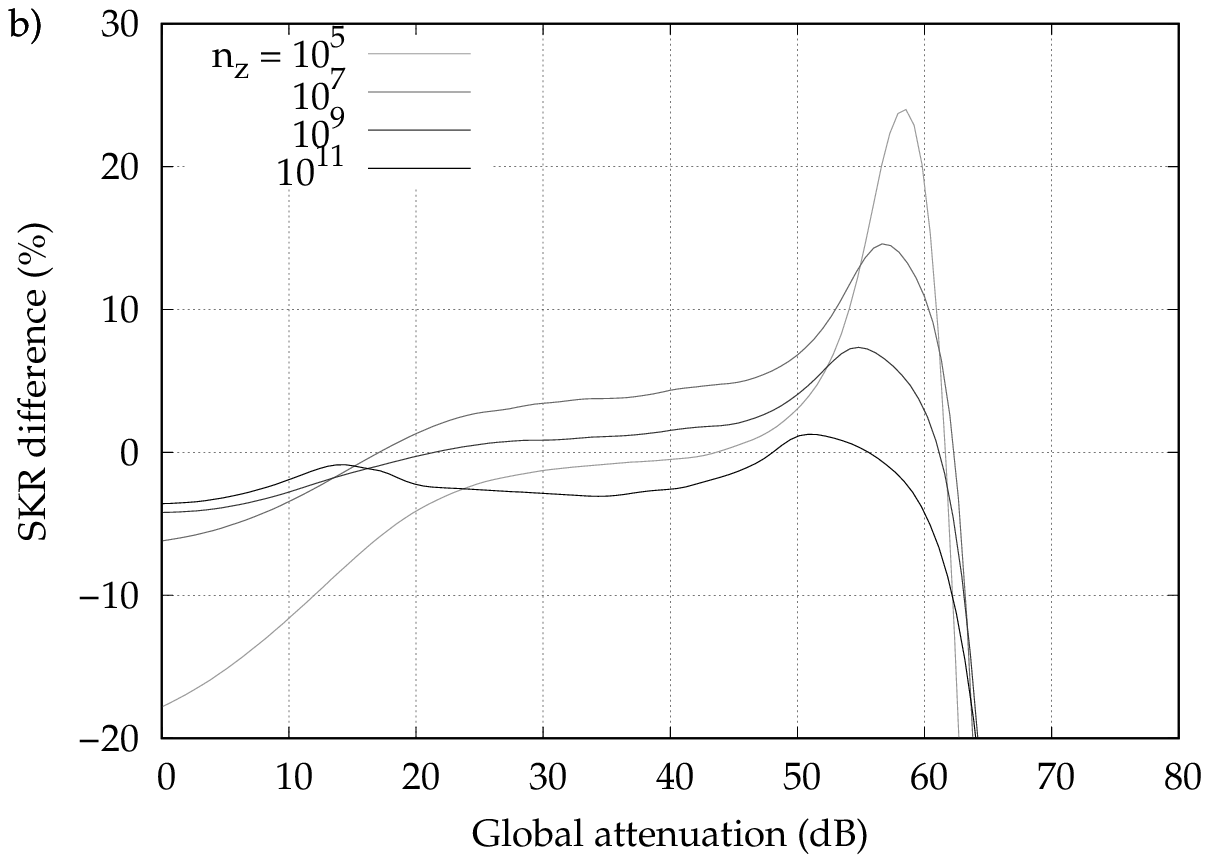}
\end{minipage}
 \caption{\label{fig:test1} (a) Comparison between different PA block sizes of the obtainable SKR considering a repetition rate of $1 \, \rm GHz$. For each block size the two protocols are shown: continuous line for the 1-decoy method and dashed line for the 2-decoy method. (b) Analysis of the percentage difference between the two protocols for different PA block sizes. (SKR\,difference $= \frac{SKR_{1D}-SKR_{2D}}{SKR_{2D}}$).}
\end{figure*}

In order to find the lower bound on this expression, another lower bound on the vacuum events $\sz{0}$ is needed. This is easily obtained by applying the decoy state analysis~\cite{Lim2014}. 

Here we continue on the same path and apply the finite-key analysis to the 1-decoy protocol (see supplementary material for more details). 
Our analysis results in a secret key length bound of the same form of Eq.~(\ref{eq_skr}).
The main difference is given by the estimation of the single-photon events.
In fact without a third intensity level the lower bound of this quantity changes to the form: 
\begin{multline}
\sz{1} \geq \sz{1}^l := \frac{\tau_1\mu_1}{\mu_2(\mu_1-\mu_2)}\left(\nzmu{2}^- -\frac{\mu_2^2}{\mu_1^2}\nzmu{1}^+\right. \\
\left.  - \frac{(\mu^2_1-\mu^2_2)}{\mu^2_1}\frac{\sz{0}^u}{\tau_0}\right).
\end{multline}
In this case, differently from the previous approach, the number of vacuum events must be upper bounded. In order to achieve this we take into account that the probability of error from a vacuum event is $1/2$. We cannot directly measure this quantity, but we can upper bound it by the total number of errors $\mzk$, for the intensity $k$. Considering the finite-key correction, we obtain the following relation (see the supplementary materials for the derivation):

\begin{multline}\label{s0}
\sz{0} \leq \sz{0}^u := 2\left(\tau_0\frac{e^k}{p_k}\left(\mzk + \sqrt{\frac{\mz}{2}\log\frac{1}{\varepsilon_2}} \right) \right.\\
\left. + \sqrt{\frac{\nz}{2}\log\frac{1}{\varepsilon_1}}\right).
\end{multline}

This is a pessimistic estimate given that the number of errors is not only due to vacuum events, i.e. dark counts and after-pulsing of the detector and counts due to parasitic light, but also by imperfections in the preparation and measurement apparatus and quantum channel de-coherence that result in a non vacuum state error.


In our simulation to maximise the SKR for a given global attenuation ($\eta$), we fix a number of parameters that depend on the characteristics of the devices and we optimize over a set of variables that can be easily tuned experimentally. For practicality, the efficiency of the detector and the internal losses of Bob's apparatus are included in the global attenuation $\eta$. The parameters considered are the probability of dark-count ($p_{\rm DC}$), the detector dead-time ($\tau_{\rm DT}$) and the alignment imperfection of the devices ($p_{\rm Err}$). 
For a given set of these parameters, we optimize the SKR over the different decoy state variables, i.e. $\mu_i$ and the associated probability $p_{\mu_i}$, and the probability to choose the \Z{} basis for Alice ($p_{Z_a}$) and Bob ($p_{Z_b}$).

The analysis in the asymptotic case was already carried out in previous works. Now, considering the finite-key scenario, the most important parameter is the number of detections in the \Z{} basis. 
This defines the privacy amplification (PA) block size $n_{Z}$ which is included in our analysis by the Hoeffding's correction.
In addition we set the secrecy and correctness parameters ($\epsilon_\text{sec}$ and $\epsilon_\text{cor}$) to the values $10^{-9}$ and $10^{-15}$ respectively, similarly to what is commonly used in literature \citep{Lim2014,korzh2015,lucamarini2013,frohlich2017}.
In \figurename{ \ref{fig:test1}(a)}, we plot the SKR for the two different approaches and for four PA block sizes.
We consider a system working at a repetition rate of 1 GHz which, as an order of magnitude, represents the source's state of the art in QKD technologies~\cite{lucamarini2013,frohlich2017}.
For the detection apparatus, we refer to recent superconducting nanowire single-photon detectors (SNSPD)~\cite{caloz2017} which have a dead-time $\tau_{\rm DT} = 100\, \rm ns $, dark-count rate (DCR) of $10\,\rm Hz$ which correspond to $p_{\rm DC} = 10^{-8}$ and an efficiency ($\eta_{det}$ around $50\%$). In the  supplementary material, we show also the analysis taking into account an InGaAs detector~\cite{Korzh2014}.
The dead-time is responsible for the saturation of the SKR at short distances, whereas the DCR at long distances is the cause of the fast drop of the SKR. 
Indeed in this regime the amount of valid detections becomes comparable to the random detector's dark counts, which raises the Quantum Bit Error Rate (QBER).
We choose a typical value $p_{\rm Err}$ of $1\%$.

\begin{figure}
\includegraphics[width = \columnwidth]{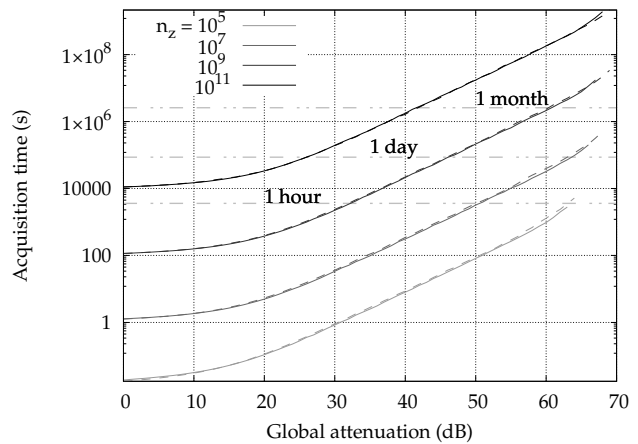}
\caption{\label{fig_time}Analysis of the time required to the QKD protocol when different block sizes are chosen. For each block size the two protocols are considered, continuous line for the 1-decoy and dashed line for the 2-decoy. For the simulations a repetition rate of 1 GHz was considered.}
\end{figure}

In this paragraph we will analyse the effect of different PA block sizes to our security analysis. 
As we see from \figurename{ \ref{fig:test1}(a)},  by increasing the block size we increase slightly the SKR as well as the maximum transmission distance. 
But, in this way, the time needed to collect the data increases proportionally to the PA block size. For this reason, in real application it is preferable to use a small PA block size. By doing this, it becomes apparent from our simulation (\figurename{ \ref{fig:test1}(b)}) that deploying 1-decoy is advantageous in most configurations. For attenuations going from 10 dB up to 60 dB it is apparent that, unless a really big ($> 10^{11}$) or really small ($< 10^5$) PA block size is applied, the simpler approach gives a higher SKR. For block sizes smaller than $10^5$ we see that for an attenuation between 40 dB and 60 dB (\figurename{ \ref{fig:test1}(b)}) the advantage of the 1-decoy protocol is still present. Moreover for small attenuation there is no practical reason to use small PA block sizes, in fact even for $n_z = 10^7$ at 40 dB the acquisition time does not exceed few minutes as presented in (\figurename{ \ref{fig_time}}).

Intuitively in an infinite-key scenario, sending the vacuum state to better estimate the $s_0$ contribution has a little positive effect on the final SKR. Indeed in this configuration even a small probability to send this intensity results in a good estimation on the vacuum events. In the case of a finite-key scenario, instead, this probability starts to be significant for reasonable block sizes. Sending a considerable amount of vacuum states diminishes the total number of detections, and consequently the SKR of the protocol. Quantitatively when the block size chosen is $\nz = 10^7$, the probability to send a vacuum state ($p_{\mu_3}$) is always greater than $10\%$ (see supplementary material); in order for this probability to go under $2\%$, the block size should be already greater than $10^{11}$.

The 2-decoy protocol turns to be useful only for either really short or really long distances. In the first case, due to the saturation of the detectors, sending vacuum states is less detrimental. However the attenuation at Bob's side (including the detector efficiency) could be high enough already at zero distance, that the detectors are no longer in the saturation regime. In the second case, even if the key exchange is possible, the results are not interesting from a practical point of view, since the SKR obtained is in the order of magnitude of 10 Hz, whereas the acquisition time starts to exceed one day. In order to give a better understanding of our thesis, we show a comparison of acquisition time and SKR for two block sizes ($\nz = 10^7$ and $\nz = 10^9$) at different distances in Tab.~\ref{tab1}. We can see that the 1-decoy protocol almost always outperforms the 2-decoy one. The only exception within the chosen attenuations appear at 64 dB; in this case however the accumulation time for a PA block starts to be impractical. 

Also other practical considerations suggest to always take the 1-decoy approach over the 2-decoy one. Having to implement only two intensity levels instead of three can give a net increase, both in terms of performances and cost efficiency of the whole system. At the same time implementing one more intensity could result in an increase of the error probability in the preparation $p_{\rm Err}$ that would decrease the SKR.

\begin{table}
\begin{tabular}{l || c | c | c | c }
Distance & ~ \tab{26\,dB\\100\,km} ~ & ~ \tab{46\,dB\\200\,km} ~ & ~ \tab{56\,dB\\250\,km} ~ & ~ \tab{64\,dB\\290\,km} ~ \\
\hline
\multicolumn{5}{c}{$n_Z = 10^7$}\\
\hline
SKR & \tabval{243\,kHz}{236\,kHz} & \tabval{2627\,Hz}{2503\,Hz} & \tabval{227\,Hz}{197\,Hz} & \tabval{11.3\,Hz}{14.1\,Hz} \\
\hline 
Time & \tabval{14\,s}{16\,s} & \tabval{20\,min}{23\,min} & \tabval{3.4\,H}{3.9\,H} & \tabval{26\,H}{31\,H}\\
\hline
\multicolumn{5}{c}{$n_Z = 10^9$}\\
\hline
SKR & \tabval{357\,kHz}{355\,kHz} & \tabval{3970\,Hz}{3881\,Hz} & \tabval{356\,Hz}{333\,Hz} & \tabval{25.5\,Hz}{30.7\,Hz} \\
\hline 
Time & \tabval{17\,min}{18\,min} & \tabval{23\,H}{24\,H} & \tabval{10\,d}{11\,d} & \tabval{67\,d}{75\,d}
\end{tabular}
\caption{\label{tab1} Comparison of SKR obtainable and time required for 1-decoy and {\color{gray2} 2-decoys} using two different PA block sizes.}
\end{table}

To conclude we presented in our work the extension of the 1-decoy protocol security to the finite-key scenario using the formalism introduced in the work of Lim et al.~\cite{Lim2014}.
By comparing the results of the finite-key effects on both 1-decoy and 2-decoy protocols we show that for practical block sizes the strategy of deploying the former protocol is advantageous. Indeed, despite the fact that we cannot measure the vacuum events directly, we achieve a higher SKR within a shorter acquisition time. We would like to stress that even if the difference between the two protocols is small, in practice they could result in a huge experimental and economical advantage.

\section*{Acknowledgements}
We would like to acknowledge Charles Ci Wen Lim for the useful discussions about the security proof. We thank the Swiss NCCR QSIT and the EUs H2020 program under the Marie Sk\l{}odowska-Curie project QCALL (GA 675662)  for financial support.

\appendix

\section{Calculation of the SKR}
\label{app_skr}

In this appendix we will describe how the different terms of Equation (1) (in the main text) are calculated from the experimental data.

This analysis follows the previous proof for a 2-decoy state protocol~\cite{Lim2014} and the general analysis for the asymptotic case done by Ma et al.~\cite{Ma2005}. 
In our protocol, we consider only a set of two intensity levels $\kappa=\{\mu_1,\mu_2\}$ where $\mu_1 > \mu_2$. 
Let us consider the case when the states are encoded in the \Z{} basis (in the basis \X{} the analysis follows in the same way) and $\sz{n}$ are the detection observed by Bob given that Alice sent an \textit{n} photon state. 
The total number of detections in the \Z{} basis are given by $\nz = \sum_{n=0}^\infty \sz{n}$.
In the asymptotic limit the number of detection with a specific intensity \textit{k} should be $\nzk^*$ where:
\begin{align}\label{eq_ndet}
\nzk^* &= \sum_{n=0}^\infty p_{k|n}\sz{n}, &\forall k \in \kappa.
\end{align} 
If we consider now a finite statistics scenario we can use Hoeffding's inequality for independent variables \cite{Hoeffding1963} by which we can bound the difference between our observed data $\nzk$ and the corresponding asymptotic case $\nzk^*$, in the following way:
\begin{equation}\label{eq_n}
|\nzk^* - \nzk| \leq \delta(\nz,\varepsilon_1),
\end{equation} 
where the former relation holds with a probability $1-2\varepsilon_1$ and $\delta(\nz,\varepsilon_1) := \sqrt{\nz\log(1/\varepsilon_1)/2}$.

The same considerations hold for the error rate estimation in a given basis.
We define the values $\vz{n}$ as the number of errors detected at Bob's side when Alice generated an $n$ photon state and $\mz = \sum_{n=0}^\infty \vz{n} $ as the total number of errors in the \Z{} basis. 
The number of errors, $\mzk^*$, for a pulse of intensity $k$, in the asymptotic case, can be expressed as:
\begin{align}
\mzk^* &= \sum_{n=0}^\infty p_{k|n}\vz{n}, &\forall k \in \kappa.
\end{align} 
Similarly to the previous case, the correction due to finite statistics is given by:
\begin{equation}\label{eq_m}
|\mzk^* - \mzk| \leq \delta(\mz,\varepsilon_2),
\end{equation} 
where the expression holds with probability $1-2\varepsilon_2$.

\subsection*{Bounds on the vacuum and single-photon events}

In order to find an analytical bound on the single-photon events, we have to define the conditional probabilities $p_{k|n}$. 
By using the Bayes' rule and by exploiting the photon distribution of a coherent state, the following expression holds:
\begin{equation}\label{eq_Bayes}
p_{k|n} = \frac{p_k}{\tau_n}p_{n|k} = \frac{p_k}{\tau_n}\frac{e^{-k}k^n}{n!},
\end{equation}
where $\tau_n = \sum_{k\in\kappa}p_ke^{-k}k^n/n!$ is the total probability to send an $n$ photon state. 
Starting from Eq.(\ref{eq_ndet}) with two different intensities, we can derive:
\begin{multline}\label{eq1} 
\frac{e^{\mu_2}\nzmu{2}}{p_{\mu_2}}-\frac{e^{\mu_1}\nzmu{1}}{p_{\mu_1}} \\ 
= \frac{(\mu_2-\mu_1)\sz{1}}{\tau_1} + \sum_{n=2}^\infty\frac{(\mu^n_{2}-\mu^n_{1})\sz{n}}{n!\tau_n} \\
\leq \frac{(\mu_2-\mu_1)s_{Z_1}}{\tau_1} +\frac{(\mu^2_2-\mu^2_1)}{\mu^2_1}\sum_{n=2}^\infty\frac{\mu_1^n\sz{n}}{n!\tau_n},
\end{multline} 
where the inequality is simply due to the fact that:
\begin{equation}
\mu^n_{2}-\mu^n_{1} = \mu^2_2\mu^{n-2}_2-\mu^2_1\mu^{n-2}_1 \leq (\mu^2_{2}-\mu^2_{1})\mu^{n-2}_{1},
\end{equation}
when $n\geq2$ and $\mu_1 > \mu_2$.
If we now consider that the sum of all the multi-photon events can be written as:
\begin{equation}
\sum_{n=2}^\infty\frac{\mu_1^n\sz{n}}{n!\tau_n} = \frac{e^{\mu_1}\nzmu{1}}{p_{\mu_1}} - \frac{\sz{0}}{\tau_0} - \mu_1\frac{\sz{1}}{\tau_1},
\end{equation}
we can rewrite the previous inequality (\ref{eq1}) as:
\begin{multline} 
\frac{e^{\mu_2}\nzmu{2}}{p_{\mu_2}}-\frac{e^{\mu_1}\nzmu{1}}{p_{\mu_1}}\leq \frac{(\mu_2-\mu_1)\sz{1}}{\tau_1} \\
+\frac{(\mu^2_2-\mu^2_1)}{\mu^2_1}\left( \frac{e^{\mu_1}\nzmu{1}}{p_{\mu_1}} - \frac{\sz{0}}{\tau_0} - \mu_1\frac{\sz{1}}{\tau_1} \right).
\end{multline}
By rearranging the terms in order to isolate the single-photon contribution $\sz{1}$ we obtain:
\begin{multline}\label{eq2} 
\sz{1} \geq \frac{\tau_1\mu_1}{\mu_2(\mu_1-\mu_2)}\left(\frac{e^{\mu_2}\nzmu{2}}{p_{\mu_2}} \right. \\
\left.-\frac{\mu_2^2}{\mu_1^2}\frac{e^{\mu_1}\nzmu{1}}{p_{\mu_1}} - \frac{(\mu^2_1-\mu^2_2)}{\mu^2_1}\frac{\sz{0}}{\tau_0}\right).
\end{multline}
 In order for this to be a lower bound on the single-photon events we have to upper bound the vacuum contribution in this expression.
 Unfortunately having only two intensity levels does not allow us to make a tight bound on this quantity.  
 The upper bound can be obtained by taking the total number of errors in one basis:
\begin{equation}
\mz = \sum_{k = \mu_1,\mu_2}p_{k|n}\sum_{n=0}^\infty\vz{n} \geq \vz{0}.
\end{equation}
Now, in order to relate this quantity to the vacuum events we use the fact that the expectation value of the errors due to vacuum ($\langle\vz{0}\rangle$) should be half of the corresponding total events~\cite{Ma2005}:

\begin{equation}\label{sz}
\frac{\langle\vz{0}\rangle}{\sz{0}} = \frac{1}{2}.
\end{equation}

\noindent This is justified by the fact that the vacuum events ($\sz{0}$) carry no information, neither for Bob nor for Eve. This means that the latter has no chance of gaining any part of the key and the former has an equal probability of having an error or a correct detection.
By taking into account the finite size effect on $\vz{0}$, we obtain:

\begin{equation}\label{vz}
\langle\vz{0}\rangle \leq \vz{0} + \delta(\sz{0},\varepsilon_1) \leq \mz + \delta(\nz,\varepsilon_1),
\end{equation}

\noindent where the second inequality holds because $\mz \geq \vz{0}$ and $\nz \geq \sz{0}$. Even if this last upper bound is not tight, it is needed since the values $\vz{0}$ and $\sz{0}$ are not directly available in the experiment, while $\mz$ and $\nz$ are. 
Finally, combining Eq.(\ref{sz}) and Eq.(\ref{vz}), we obtain the upper bound on $\sz{0}$:

\begin{equation}
\sz{0} \leq \sz{0}^u := 2\left(\mz + \delta(\nz,\varepsilon_1)\right).
\end{equation}

Another analogous way to obtain an upper bound on the vacuum events is to consider only the errors relative to one intensity. In this case we have the relation:

\begin{multline}
\mzk^* = \sum_{n=0}^\infty p_{k|n}\vz{n} = \sum_{n=0}^\infty \frac{p_k}{\tau_n}\frac{e^{-k}k^n}{n!}\vz{n} \\
\geq \frac{p_k}{\tau_o}e^{-k}\vz{0} = \frac{p_k}{\tau_o}e^{-k}\vz{0}.
\end{multline}

As said in the main text the second approach was chosen, which proved to give the best SKR. In this scenario by taking into account the finite-key statistic the upper bound becomes:

\begin{multline}
\sz{0} \leq \sz{0}^u := 2\left(\tau_0\frac{e^k}{p_k}\left(\mzk + \delta(\mz,\varepsilon_1) \right) \right.\\
\left. + \delta(\nz,\varepsilon_1)\vphantom{\frac{n^n}{n_n}}\right).
\end{multline}

By implementing this last result in the inequality (\ref{eq2}) and applying the finite-key corrections to it, we obtain:
\begin{multline}
\sz{1} \geq \sz{1}^l := \frac{\tau_1\mu_1}{\mu_2(\mu_1-\mu_2)}\left(\nzmu{2}^-\right. \\
\left. -\frac{\mu_2^2}{\mu_1^2}\nzmu{1}^+ - \frac{(\mu^2_1-\mu^2_2)}{\mu^2_1}\frac{\sz{0}^u}{\tau_0}\right),
\end{multline}
where we defined:
\begin{align}
\nzk^\pm &:=   \left(\nzk \pm \delta(\nz,\varepsilon_1)\right), \forall k \in \kappa.
\end{align}
The lower bound on the vacuum events, in the finite-key scenario, is given by the formula \cite{Lim2014}:
\begin{equation}
\sz{0} \geq \sz{0}^l := \frac{\tau_0}{\mu_1-\mu_2}\left(\mu_1\nzmu{2}^- -\mu_2\nzmu{1}^+\right).
\end{equation}

\subsection*{Phase error rate}

In order to estimate the phase error in the \Z{} basis, the following formula can be used ~\cite{fung2010}:
\begin{equation}
\phiz := \frac{c_{\mathsf{Z},1}}{\sz{1}} \leq \frac{\vx{1}}{\sx{1}} + \gamma\left(\varepsilon_\text{sec},\frac{\vx{1}}{\sx{1}},\sz{1},\sx{1}\right),
\end{equation}
where:
\begin{multline}
\gamma\left(a,b,c,d\right)\\
 = \sqrt{\frac{(c+d)(1-b)b}{cd\log2}\log_2\left(\frac{c+d}{cd(1-b)b}\frac{21^2}{a^2}\right)}.
\end{multline}

Now by using the same result as in \cite{Lim2014} we can upper bound the number of bit errors in the \X{} basis due to single-photons by the analytic formula that follows:
\begin{equation}
\vx{1} \leq \vx{1}^u = \frac{\tau_1}{\mu_1-\mu_2}\left(\mxmu{1}^+ -\mxmu{2}^-\right).
\end{equation}
With this we can also upper bound the phase error rate in the \Z{} basis by the formula:
\begin{equation}
\phiz \leq \phi^{u}_x := \frac{\vx{1}^{u}}{\sx{1}^l} + \gamma\left(\varepsilon_{sec},\frac{\vx{1}^u}{\sx{1}^l},\sz{1}^l,\sx{1}^l\right).
\end{equation}

We have now all the terms needed to estimate the secret key length.

\subsection*{Secret Key Length parameters $a$ and $b$}

In case of the 2-decoy approach the complete analysis was already carried out by Lim et al.~\cite{Lim2014}. Their security analysis resulted in the specific values of $a = 6$, $b = 21$. 

In our work we followed the same security analysis approach. The only difference is given by the definition of the security parameter $\epsilon_\text{sec}$. In our analysis in fact this parameter has the form:

\begin{equation}
\epsilon_\text{sec} = 2\left[\alpha_1 + 2\alpha_2 + \alpha_3\right] + \nu + 6\epsilon_1 + 4\epsilon_2,
\end{equation}

where $\alpha_1, \alpha_2, \alpha_3$ and $\nu$ are error terms carried out in the security analysis~\cite{Lim2014}. The coefficients in front of the terms $\epsilon_1$ and $\epsilon_2$ are equal to the number of times the concentration inequalities, Eq.(\ref{eq_n}) and Eq.(\ref{eq_m}) respectively, were implemented in the secret key length formula. By setting all error terms by a common value $\epsilon$ we obtain $\epsilon_\text{sec} = 19\epsilon$ which implemented in the security proof permit us to find a secret key length formula for the 1-decoy approach of the form:

\begin{align}\label{eq_skr}
l \leq &  \sz{0}^l + \sz{1}^l(1-h(\phiz^u)) - \lambda_\text{EC} \nonumber\\
& - 6\log_2(19/\epsilon_\text{sec}) - \log_2(2/\epsilon_\text{cor}).
\end{align}

\begin{figure}
\includegraphics[width = \columnwidth]{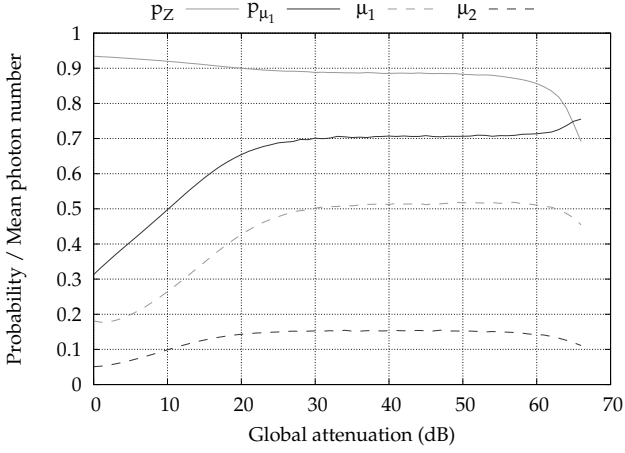}
\caption{\label{par_1}Optimization variables evolution over increasing attenuation for the 1-decoy protocol.}
\end{figure}

\begin{figure}
\includegraphics[width = \columnwidth]{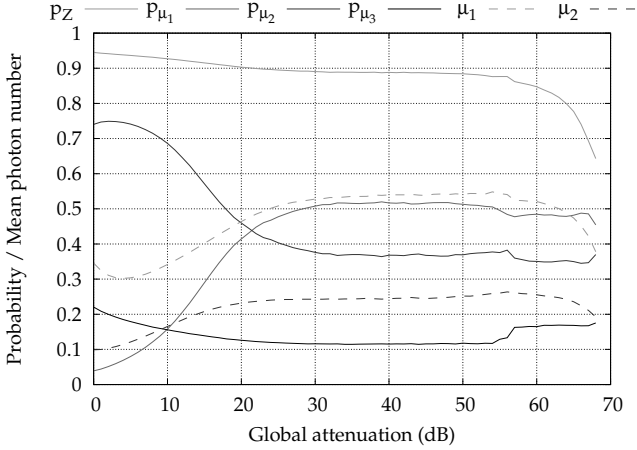}
\caption{\label{par_2}Optimization variables evolution over increasing attenuation for the 2-decoy protocol. Note that $\mu_3 = 10^{-6}$ for all attenuations.}
\end{figure}

\begin{figure*}
\centering
\begin{minipage}{.5\textwidth}
\includegraphics[width=1\linewidth]{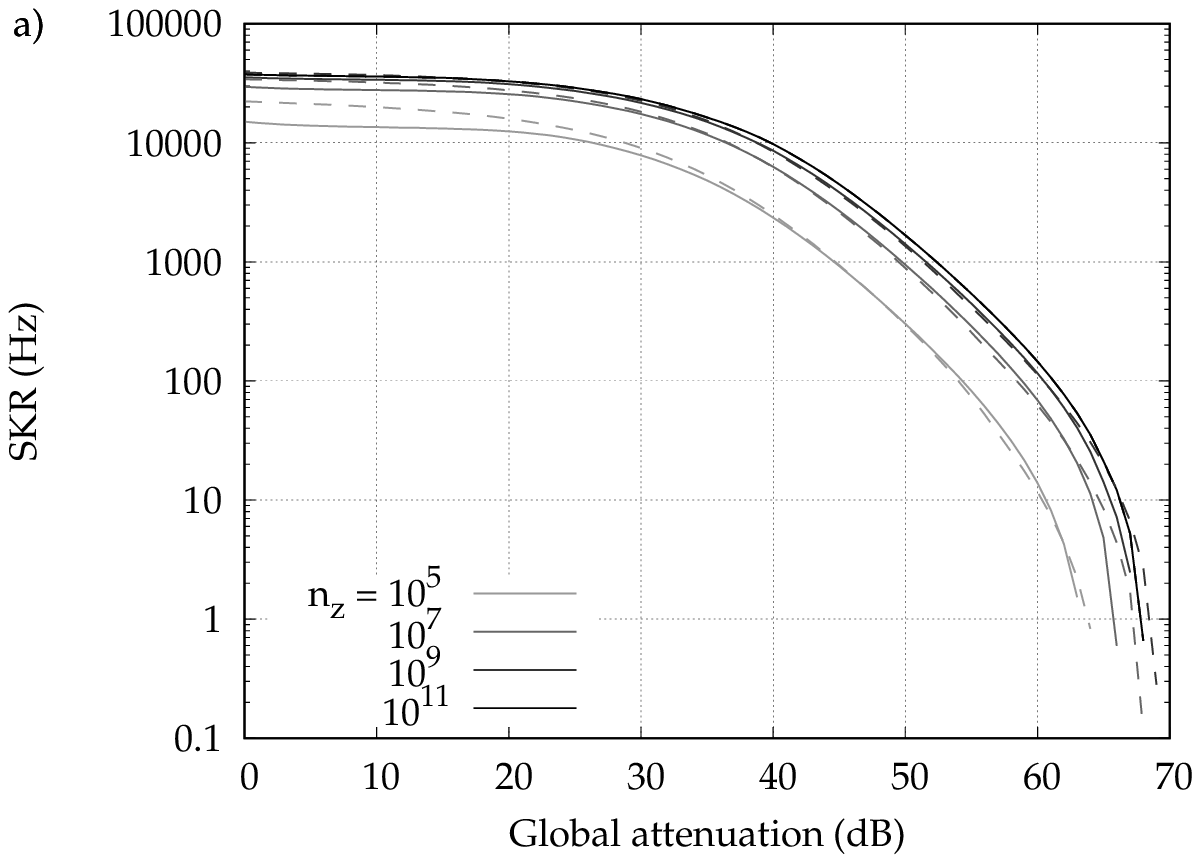}
\end{minipage}%
\begin{minipage}{.5\textwidth}
  \centering
  \includegraphics[width=1\linewidth]{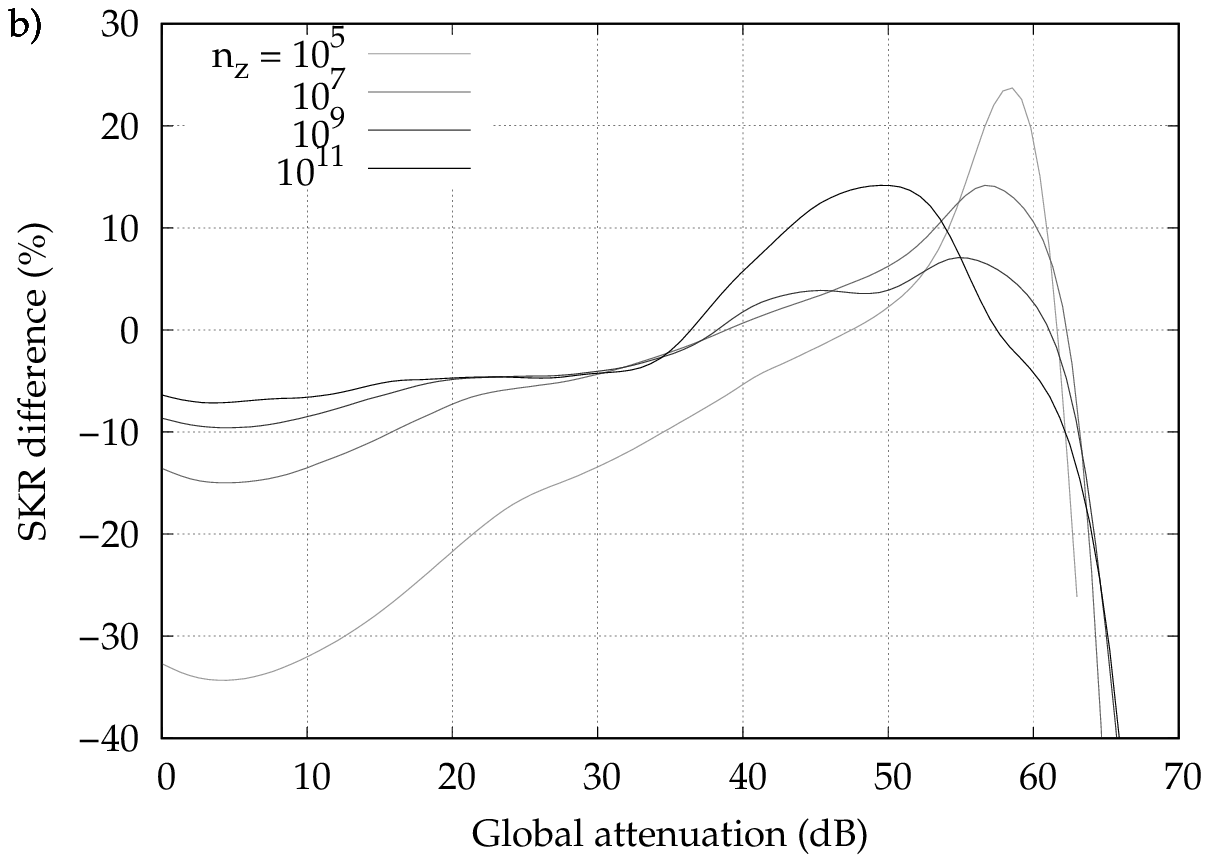}
\end{minipage}
 \caption{\label{NFAT} a) Analysis of the SKR for different PA block sizes. The continuous line represents the 1-decoy protocol and the dashed line the 2-decoy one. b) Comparison of the difference between the SKR of the 1-decoy and 2-decoy protocols for different PA block sizes.}
\end{figure*}

\section{Detection and Error Simulation}

In the simulation, as already mentioned in the text, the total number of detection in the \Z{} basis is fixed ($\nz$). In the following section we show how we simulate the different detections and error events in the \Z{} basis; the analysis for the \X{} basis is analogous.
In order to simulate the number of detections, $\nzmu{i}$, due to a certain intensity $\mu_i$ we calculate the corresponding fraction of $\nz$ as:

\begin{equation}\label{nzmu}
\nzmu{i} = \nz \frac{P_{\mathsf{Z},det,\mu_i}}{P_{\mathsf{Z},det,tot}}.
\end{equation}

\noindent In this expression $P_{\mathsf{Z},det,\mu_i}$ is the detection probability in the \Z{} basis due to a pulse of intensity $\mu_i$ after an attenuation of $\eta$, and $P_{\mathsf{Z},det,tot}$ is the sum of these probabilities over all possible intensities. This probability can be easily found to be equal to: 

\begin{equation}
P_{\mathsf{Z},det,\mu_i} = c_{DT} P_{\mathsf{Z}} P_{\mu_i} \left(\left(1 - e^{\left(- \mu_i \eta\right)}\right) + P_{DC}\right).
\end{equation} 

\noindent where $P_{\mathsf{Z}}$ is the probability that both Alice and Bob chose the \Z{} basis, $P_{\mu_i}$ is the probability to send the decoy $\mu_i$ and $c_{DT}$ is the correction factor due to the dead time ($t_{DT}$) of the detectors. We model this contribution after the expression:

\begin{equation}
c_{dt} = \frac{1}{1 + R P_{\mathsf{Z},det,tot} t_{DT}},
\end{equation}

\noindent where $R$ is the repetition rate of the source.

The error probability is then given by the formula:

\begin{equation}
P_{\mathsf{Z},err,\mu_i} = c_{dt} P_{\mathsf{Z}} P_{\mu_i} \left(\left(1 - e^{\left(- \mu_i \eta\right)}\right)P_{Err} + \frac{P_{DC}}{2}\right),
\end{equation} 

\noindent where $P_{Err}$ and $P_{DC}$ are, as already stated in the manuscript, the probabilities of error due to the misalignment of the set-up and due to the detectors' dark counts, respectively.  

In order to calculate the QBER on the \Z{} basis, one can just evaluate the ratio between the total probability of error and the total probability of detection:

\begin{equation}
\textrm{QBER} = \frac{\mz{}}{\nz{}} = \frac{P_{\mathsf{Z},err,tot}}{P_{\mathsf{Z},det,tot}}.
\end{equation}

To find the number of errors due to a pulse of intensity $\mu_i$ we proceed similarly to Eq.(\ref{nzmu}):

\begin{equation}
\mzmu{i} = \nz \frac{P_{\mathsf{Z},err,\mu_i}}{P_{\mathsf{Z},det,tot}}.
\end{equation}

Hence, the number of pulses that need to be sent ($N_{tot}$) in order to obtain a block size of $\nz$ is easily found to be equal to the following expression:

\begin{equation}
N_{tot} = \frac{\nz}{P_{\mathsf{Z},det,tot}}.
\end{equation}

Finally the \textrm{SKR} is found by taking the ratio between the secret key length Eq.(\ref{eq_skr}) and the total number of pulse sent and multiplying it by the repetition rate of the source ($R$):

\begin{equation}
\textrm{SKR} = \frac{l}{N_{tot}}R.
\end{equation} 

\section{Simulation Variables}
\label{app_var}

We show, for completeness, the values of the different variables chosen in the optimization process for both protocols in case of a chosen block size of $10^7$. In \figurename{~\ref{par_1}} and \figurename{~\ref{par_2}} are shown the probabilities $p_Z = p_{Z_a} = p_{Z_b}$ and $p_{\mu_i}$ and the mean photon number $\mu_i$ of the signal and decoy state as a function of the attenuation. The intensities are small for low attenuations, since in this regime the detectors are saturated. For higher attenuations they then increase and remain constant at their optimal values. The probability to choose the \Z{} basis for Alice and Bob cannot be taken equal to one due to the finite block size: a part of the pulses sent must be used to estimate the phase error rate using the complementary \X{} basis. In the analysis of the 2-decoy protocol (\figurename{~\ref{par_2}}) we want to outline that the probability to send the vacuum state $\mu_3$ is non negligible for all attenuations: even at its minimum value it remains always greater than $11\%$.

\section{Simulation with I\lowercase{n}G\lowercase{a}A\lowercase{s} Detectors}
\label{app_InGaAs}
In this section we show the behaviour of the SKR obtained by modelling an InGaAs detector~\cite{Korzh2014} with a DCR of $1\, \rm Hz$ and a dead time of $20\, \rm \mu s$.

The higher dead time of these detectors, compared to the SNSPDs, increases the attenuation interval in which the SKR is mainly limited by the saturation of the detectors (\figurename{ \ref{NFAT}a}). As a direct consequence, the attenuation range in which the 2-decoy protocol outperforms the 1-decoy one also increases (\figurename{ \ref{NFAT}b}). This is due to the advantages of the 2-decoy protocol in this regime explained in the main text. Nevertheless, for attenuations between 30 dB and 70 dB, the 1-decoy protocol results in the best SKR for practical PA block sizes.

\bibliography{Bib_1vs2decoy}

\begin{thebibliography}{21}%
\makeatletter
\providecommand \@ifxundefined [1]{%
 \@ifx{#1\undefined}
}%
\providecommand \@ifnum [1]{%
 \ifnum #1\expandafter \@firstoftwo
 \else \expandafter \@secondoftwo
 \fi
}%
\providecommand \@ifx [1]{%
 \ifx #1\expandafter \@firstoftwo
 \else \expandafter \@secondoftwo
 \fi
}%
\providecommand \natexlab [1]{#1}%
\providecommand \enquote  [1]{``#1''}%
\providecommand \bibnamefont  [1]{#1}%
\providecommand \bibfnamefont [1]{#1}%
\providecommand \citenamefont [1]{#1}%
\providecommand \href@noop [0]{\@secondoftwo}%
\providecommand \href [0]{\begingroup \@sanitize@url \@href}%
\providecommand \@href[1]{\@@startlink{#1}\@@href}%
\providecommand \@@href[1]{\endgroup#1\@@endlink}%
\providecommand \@sanitize@url [0]{\catcode `\\12\catcode `\$12\catcode
  `\&12\catcode `\#12\catcode `\^12\catcode `\_12\catcode `\%12\relax}%
\providecommand \@@startlink[1]{}%
\providecommand \@@endlink[0]{}%
\providecommand \url  [0]{\begingroup\@sanitize@url \@url }%
\providecommand \@url [1]{\endgroup\@href {#1}{\urlprefix }}%
\providecommand \urlprefix  [0]{URL }%
\providecommand \Eprint [0]{\href }%
\providecommand \doibase [0]{http://dx.doi.org/}%
\providecommand \selectlanguage [0]{\@gobble}%
\providecommand \bibinfo  [0]{\@secondoftwo}%
\providecommand \bibfield  [0]{\@secondoftwo}%
\providecommand \translation [1]{[#1]}%
\providecommand \BibitemOpen [0]{}%
\providecommand \bibitemStop [0]{}%
\providecommand \bibitemNoStop [0]{.\EOS\space}%
\providecommand \EOS [0]{\spacefactor3000\relax}%
\providecommand \BibitemShut  [1]{\csname bibitem#1\endcsname}%
\let\auto@bib@innerbib\@empty
\bibitem [{\citenamefont {Bennett}\ and\ \citenamefont
  {Brassard}(1984)}]{Bennett1984}%
  \BibitemOpen
  \bibfield  {author} {\bibinfo {author} {\bibfnamefont {C.~H.}\ \bibnamefont
  {Bennett}}\ and\ \bibinfo {author} {\bibfnamefont {G.}~\bibnamefont
  {Brassard}},\ }in\ \href@noop {} {\emph {\bibinfo {booktitle} {International
  Conference on Computers, Systems \& Signal Processing, Bangalore, India, Dec
  9-12, 1984}}}\ (\bibinfo {year} {1984})\ pp.\ \bibinfo {pages}
  {175--179}\BibitemShut {NoStop}%
\bibitem [{\citenamefont {Gisin}\ \emph {et~al.}(2002)\citenamefont {Gisin},
  \citenamefont {Ribordy}, \citenamefont {Tittel},\ and\ \citenamefont
  {Zbinden}}]{Gisin2002}%
  \BibitemOpen
  \bibfield  {author} {\bibinfo {author} {\bibfnamefont {N.}~\bibnamefont
  {Gisin}}, \bibinfo {author} {\bibfnamefont {G.}~\bibnamefont {Ribordy}},
  \bibinfo {author} {\bibfnamefont {W.}~\bibnamefont {Tittel}}, \ and\ \bibinfo
  {author} {\bibfnamefont {H.}~\bibnamefont {Zbinden}},\ }\href {\doibase
  10.1103/RevModPhys.74.145} {\bibfield  {journal} {\bibinfo  {journal} {Rev.
  Mod. Phys.}\ }\textbf {\bibinfo {volume} {74}},\ \bibinfo {pages} {145}
  (\bibinfo {year} {2002})}\BibitemShut {NoStop}%
\bibitem [{\citenamefont {Scarani}\ \emph {et~al.}(2009)\citenamefont
  {Scarani}, \citenamefont {Bechmann-Pasquinucci}, \citenamefont {Cerf},
  \citenamefont {Du{\v{s}}ek}, \citenamefont {L{\"u}tkenhaus},\ and\
  \citenamefont {Peev}}]{scarani2009}%
  \BibitemOpen
  \bibfield  {author} {\bibinfo {author} {\bibfnamefont {V.}~\bibnamefont
  {Scarani}}, \bibinfo {author} {\bibfnamefont {H.}~\bibnamefont
  {Bechmann-Pasquinucci}}, \bibinfo {author} {\bibfnamefont {N.~J.}\
  \bibnamefont {Cerf}}, \bibinfo {author} {\bibfnamefont {M.}~\bibnamefont
  {Du{\v{s}}ek}}, \bibinfo {author} {\bibfnamefont {N.}~\bibnamefont
  {L{\"u}tkenhaus}}, \ and\ \bibinfo {author} {\bibfnamefont {M.}~\bibnamefont
  {Peev}},\ }\href {\doibase 10.1103/RevModPhys.81.1301} {\bibfield  {journal}
  {\bibinfo  {journal} {Reviews of modern physics}\ }\textbf {\bibinfo {volume}
  {81}},\ \bibinfo {pages} {1301} (\bibinfo {year} {2009})}\BibitemShut
  {NoStop}%
\bibitem [{\citenamefont {Huttner}\ \emph {et~al.}(1995)\citenamefont
  {Huttner}, \citenamefont {Imoto}, \citenamefont {Gisin},\ and\ \citenamefont
  {Mor}}]{huttner1995}%
  \BibitemOpen
  \bibfield  {author} {\bibinfo {author} {\bibfnamefont {B.}~\bibnamefont
  {Huttner}}, \bibinfo {author} {\bibfnamefont {N.}~\bibnamefont {Imoto}},
  \bibinfo {author} {\bibfnamefont {N.}~\bibnamefont {Gisin}}, \ and\ \bibinfo
  {author} {\bibfnamefont {T.}~\bibnamefont {Mor}},\ }\href {\doibase
  10.1103/PhysRevA.51.1863} {\bibfield  {journal} {\bibinfo  {journal}
  {Physical Review A}\ }\textbf {\bibinfo {volume} {51}},\ \bibinfo {pages}
  {1863} (\bibinfo {year} {1995})}\BibitemShut {NoStop}%
\bibitem [{\citenamefont {Brassard}\ \emph {et~al.}(2000)\citenamefont
  {Brassard}, \citenamefont {L{\"u}tkenhaus}, \citenamefont {Mor},\ and\
  \citenamefont {Sanders}}]{brassard2000}%
  \BibitemOpen
  \bibfield  {author} {\bibinfo {author} {\bibfnamefont {G.}~\bibnamefont
  {Brassard}}, \bibinfo {author} {\bibfnamefont {N.}~\bibnamefont
  {L{\"u}tkenhaus}}, \bibinfo {author} {\bibfnamefont {T.}~\bibnamefont {Mor}},
  \ and\ \bibinfo {author} {\bibfnamefont {B.~C.}\ \bibnamefont {Sanders}},\
  }\href {\doibase 10.1103/PhysRevLett.85.1330} {\bibfield  {journal} {\bibinfo
   {journal} {Physical Review Letters}\ }\textbf {\bibinfo {volume} {85}},\
  \bibinfo {pages} {1330} (\bibinfo {year} {2000})}\BibitemShut {NoStop}%
\bibitem [{\citenamefont {Scarani}\ \emph {et~al.}(2004)\citenamefont
  {Scarani}, \citenamefont {Ac\'{\i}n}, \citenamefont {Ribordy},\ and\
  \citenamefont {Gisin}}]{Scarani2004}%
  \BibitemOpen
  \bibfield  {author} {\bibinfo {author} {\bibfnamefont {V.}~\bibnamefont
  {Scarani}}, \bibinfo {author} {\bibfnamefont {A.}~\bibnamefont {Ac\'{\i}n}},
  \bibinfo {author} {\bibfnamefont {G.}~\bibnamefont {Ribordy}}, \ and\
  \bibinfo {author} {\bibfnamefont {N.}~\bibnamefont {Gisin}},\ }\href
  {\doibase 10.1103/PhysRevLett.92.057901} {\bibfield  {journal} {\bibinfo
  {journal} {Phys. Rev. Lett.}\ }\textbf {\bibinfo {volume} {92}},\ \bibinfo
  {pages} {057901} (\bibinfo {year} {2004})}\BibitemShut {NoStop}%
\bibitem [{\citenamefont {Inoue}\ \emph {et~al.}(2002)\citenamefont {Inoue},
  \citenamefont {Waks},\ and\ \citenamefont {Yamamoto}}]{Inoue2002}%
  \BibitemOpen
  \bibfield  {author} {\bibinfo {author} {\bibfnamefont {K.}~\bibnamefont
  {Inoue}}, \bibinfo {author} {\bibfnamefont {E.}~\bibnamefont {Waks}}, \ and\
  \bibinfo {author} {\bibfnamefont {Y.}~\bibnamefont {Yamamoto}},\ }\href
  {\doibase 10.1103/PhysRevLett.89.037902} {\bibfield  {journal} {\bibinfo
  {journal} {Phys. Rev. Lett.}\ }\textbf {\bibinfo {volume} {89}},\ \bibinfo
  {pages} {037902} (\bibinfo {year} {2002})}\BibitemShut {NoStop}%
\bibitem [{\citenamefont {Stucki}\ \emph {et~al.}(2005)\citenamefont {Stucki},
  \citenamefont {Brunner}, \citenamefont {Gisin}, \citenamefont {Scarani},\
  and\ \citenamefont {Zbinden}}]{Stucki2005}%
  \BibitemOpen
  \bibfield  {author} {\bibinfo {author} {\bibfnamefont {D.}~\bibnamefont
  {Stucki}}, \bibinfo {author} {\bibfnamefont {N.}~\bibnamefont {Brunner}},
  \bibinfo {author} {\bibfnamefont {N.}~\bibnamefont {Gisin}}, \bibinfo
  {author} {\bibfnamefont {V.}~\bibnamefont {Scarani}}, \ and\ \bibinfo
  {author} {\bibfnamefont {H.}~\bibnamefont {Zbinden}},\ }\href {\doibase
  10.1063/1.2126792} {\bibfield  {journal} {\bibinfo  {journal} {Appl. Phys.
  Lett.}\ }\textbf {\bibinfo {volume} {87}},\ \bibinfo {pages} {194108}
  (\bibinfo {year} {2005})}\BibitemShut {NoStop}%
\bibitem [{\citenamefont {Hwang}(2003)}]{Hwang2003}%
  \BibitemOpen
  \bibfield  {author} {\bibinfo {author} {\bibfnamefont {W.-Y.}\ \bibnamefont
  {Hwang}},\ }\href {\doibase 10.1103/physrevlett.91.057901} {\bibfield
  {journal} {\bibinfo  {journal} {Phys. Rev. Lett.}\ }\textbf {\bibinfo
  {volume} {91}} (\bibinfo {year} {2003}),\
  10.1103/physrevlett.91.057901}\BibitemShut {NoStop}%
\bibitem [{\citenamefont {Lo}\ \emph {et~al.}(2005)\citenamefont {Lo},
  \citenamefont {Ma},\ and\ \citenamefont {Chen}}]{Lo2004}%
  \BibitemOpen
  \bibfield  {author} {\bibinfo {author} {\bibfnamefont {H.-K.}\ \bibnamefont
  {Lo}}, \bibinfo {author} {\bibfnamefont {X.}~\bibnamefont {Ma}}, \ and\
  \bibinfo {author} {\bibfnamefont {K.}~\bibnamefont {Chen}},\ }\href {\doibase
  10.1103/PhysRevLett.94.230504} {\bibfield  {journal} {\bibinfo  {journal}
  {Phys. Rev. Lett.}\ }\textbf {\bibinfo {volume} {94}},\ \bibinfo {pages}
  {230504} (\bibinfo {year} {2005})},\ \Eprint {http://arxiv.org/abs/0411004}
  {arXiv:0411004 [quant-ph]} \BibitemShut {NoStop}%
\bibitem [{\citenamefont {Wang}(2005)}]{wang2005}%
  \BibitemOpen
  \bibfield  {author} {\bibinfo {author} {\bibfnamefont {X.-B.}\ \bibnamefont
  {Wang}},\ }\href {\doibase 10.1103/PhysRevLett.94.230503} {\bibfield
  {journal} {\bibinfo  {journal} {Physical review letters}\ }\textbf {\bibinfo
  {volume} {94}},\ \bibinfo {pages} {230503} (\bibinfo {year}
  {2005})}\BibitemShut {NoStop}%
\bibitem [{\citenamefont {Ma}\ \emph {et~al.}(2005)\citenamefont {Ma},
  \citenamefont {Qi}, \citenamefont {Zhao},\ and\ \citenamefont {Lo}}]{Ma2005}%
  \BibitemOpen
  \bibfield  {author} {\bibinfo {author} {\bibfnamefont {X.}~\bibnamefont
  {Ma}}, \bibinfo {author} {\bibfnamefont {B.}~\bibnamefont {Qi}}, \bibinfo
  {author} {\bibfnamefont {Y.}~\bibnamefont {Zhao}}, \ and\ \bibinfo {author}
  {\bibfnamefont {H.-K.}\ \bibnamefont {Lo}},\ }\href {\doibase
  10.1103/PhysRevA.72.012326} {\bibfield  {journal} {\bibinfo  {journal} {Phys.
  Rev. A}\ }\textbf {\bibinfo {volume} {72}},\ \bibinfo {pages} {012326}
  (\bibinfo {year} {2005})}\BibitemShut {NoStop}%
\bibitem [{\citenamefont {Hayashi}\ and\ \citenamefont
  {Nakayama}(2014)}]{hayashi2014}%
  \BibitemOpen
  \bibfield  {author} {\bibinfo {author} {\bibfnamefont {M.}~\bibnamefont
  {Hayashi}}\ and\ \bibinfo {author} {\bibfnamefont {R.}~\bibnamefont
  {Nakayama}},\ }\href {\doibase 10.1088/1367-2630/16/6/063009} {\bibfield
  {journal} {\bibinfo  {journal} {New Journal of Physics}\ }\textbf {\bibinfo
  {volume} {16}},\ \bibinfo {pages} {063009} (\bibinfo {year}
  {2014})}\BibitemShut {NoStop}%
\bibitem [{\citenamefont {Lim}\ \emph {et~al.}(2014)\citenamefont {Lim},
  \citenamefont {Curty}, \citenamefont {Walenta}, \citenamefont {Xu},\ and\
  \citenamefont {Zbinden}}]{Lim2014}%
  \BibitemOpen
  \bibfield  {author} {\bibinfo {author} {\bibfnamefont {C.~C.~W.}\
  \bibnamefont {Lim}}, \bibinfo {author} {\bibfnamefont {M.}~\bibnamefont
  {Curty}}, \bibinfo {author} {\bibfnamefont {N.}~\bibnamefont {Walenta}},
  \bibinfo {author} {\bibfnamefont {F.}~\bibnamefont {Xu}}, \ and\ \bibinfo
  {author} {\bibfnamefont {H.}~\bibnamefont {Zbinden}},\ }\href {\doibase
  10.1103/PhysRevA.89.022307} {\bibfield  {journal} {\bibinfo  {journal} {Phys.
  Rev. A}\ }\textbf {\bibinfo {volume} {89}} (\bibinfo {year} {2014}),\
  10.1103/PhysRevA.89.022307}\BibitemShut {NoStop}%
\bibitem [{\citenamefont {Hoeffding}(1963)}]{Hoeffding1963}%
  \BibitemOpen
  \bibfield  {author} {\bibinfo {author} {\bibfnamefont {W.}~\bibnamefont
  {Hoeffding}},\ }\href {\doibase 10.1080/01621459.1963.10500830} {\bibfield
  {journal} {\bibinfo  {journal} {J. Am. Stat. Assoc.}\ }\textbf {\bibinfo
  {volume} {58}},\ \bibinfo {pages} {13} (\bibinfo {year} {1963})}\BibitemShut
  {NoStop}%
\bibitem [{\citenamefont {Korzh}\ \emph {et~al.}(2015)\citenamefont {Korzh},
  \citenamefont {Lim}, \citenamefont {Houlmann}, \citenamefont {Gisin},
  \citenamefont {Li}, \citenamefont {Nolan}, \citenamefont {Sanguinetti},
  \citenamefont {Thew},\ and\ \citenamefont {Zbinden}}]{korzh2015}%
  \BibitemOpen
  \bibfield  {author} {\bibinfo {author} {\bibfnamefont {B.}~\bibnamefont
  {Korzh}}, \bibinfo {author} {\bibfnamefont {C.~C.~W.}\ \bibnamefont {Lim}},
  \bibinfo {author} {\bibfnamefont {R.}~\bibnamefont {Houlmann}}, \bibinfo
  {author} {\bibfnamefont {N.}~\bibnamefont {Gisin}}, \bibinfo {author}
  {\bibfnamefont {M.~J.}\ \bibnamefont {Li}}, \bibinfo {author} {\bibfnamefont
  {D.}~\bibnamefont {Nolan}}, \bibinfo {author} {\bibfnamefont
  {B.}~\bibnamefont {Sanguinetti}}, \bibinfo {author} {\bibfnamefont
  {R.}~\bibnamefont {Thew}}, \ and\ \bibinfo {author} {\bibfnamefont
  {H.}~\bibnamefont {Zbinden}},\ }\href {\doibase 10.1038/nphoton.2014.327}
  {\bibfield  {journal} {\bibinfo  {journal} {Nature Photonics}\ }\textbf
  {\bibinfo {volume} {9}},\ \bibinfo {pages} {163} (\bibinfo {year}
  {2015})}\BibitemShut {NoStop}%
\bibitem [{\citenamefont {Lucamarini}\ \emph {et~al.}(2013)\citenamefont
  {Lucamarini}, \citenamefont {Patel}, \citenamefont {Dynes}, \citenamefont
  {Fr{\"o}hlich}, \citenamefont {Sharpe}, \citenamefont {Dixon}, \citenamefont
  {Yuan}, \citenamefont {Penty},\ and\ \citenamefont
  {Shields}}]{lucamarini2013}%
  \BibitemOpen
  \bibfield  {author} {\bibinfo {author} {\bibfnamefont {M.}~\bibnamefont
  {Lucamarini}}, \bibinfo {author} {\bibfnamefont {K.}~\bibnamefont {Patel}},
  \bibinfo {author} {\bibfnamefont {J.}~\bibnamefont {Dynes}}, \bibinfo
  {author} {\bibfnamefont {B.}~\bibnamefont {Fr{\"o}hlich}}, \bibinfo {author}
  {\bibfnamefont {A.}~\bibnamefont {Sharpe}}, \bibinfo {author} {\bibfnamefont
  {A.}~\bibnamefont {Dixon}}, \bibinfo {author} {\bibfnamefont
  {Z.}~\bibnamefont {Yuan}}, \bibinfo {author} {\bibfnamefont {R.}~\bibnamefont
  {Penty}}, \ and\ \bibinfo {author} {\bibfnamefont {A.}~\bibnamefont
  {Shields}},\ }\href {\doibase 10.1364/OE.21.024550} {\bibfield  {journal}
  {\bibinfo  {journal} {Optics express}\ }\textbf {\bibinfo {volume} {21}},\
  \bibinfo {pages} {24550} (\bibinfo {year} {2013})}\BibitemShut {NoStop}%
\bibitem [{\citenamefont {Fr{\"o}hlich}\ \emph {et~al.}(2017)\citenamefont
  {Fr{\"o}hlich}, \citenamefont {Lucamarini}, \citenamefont {Dynes},
  \citenamefont {Comandar}, \citenamefont {Tam}, \citenamefont {Plews},
  \citenamefont {Sharpe}, \citenamefont {Yuan},\ and\ \citenamefont
  {Shields}}]{frohlich2017}%
  \BibitemOpen
  \bibfield  {author} {\bibinfo {author} {\bibfnamefont {B.}~\bibnamefont
  {Fr{\"o}hlich}}, \bibinfo {author} {\bibfnamefont {M.}~\bibnamefont
  {Lucamarini}}, \bibinfo {author} {\bibfnamefont {J.~F.}\ \bibnamefont
  {Dynes}}, \bibinfo {author} {\bibfnamefont {L.~C.}\ \bibnamefont {Comandar}},
  \bibinfo {author} {\bibfnamefont {W.~W.-S.}\ \bibnamefont {Tam}}, \bibinfo
  {author} {\bibfnamefont {A.}~\bibnamefont {Plews}}, \bibinfo {author}
  {\bibfnamefont {A.~W.}\ \bibnamefont {Sharpe}}, \bibinfo {author}
  {\bibfnamefont {Z.}~\bibnamefont {Yuan}}, \ and\ \bibinfo {author}
  {\bibfnamefont {A.~J.}\ \bibnamefont {Shields}},\ }\href {\doibase
  10.1364/OPTICA.4.000163} {\bibfield  {journal} {\bibinfo  {journal} {Optica}\
  }\textbf {\bibinfo {volume} {4}},\ \bibinfo {pages} {163} (\bibinfo {year}
  {2017})}\BibitemShut {NoStop}%
\bibitem [{\citenamefont {Caloz}\ \emph {et~al.}(2017)\citenamefont {Caloz},
  \citenamefont {Korzh}, \citenamefont {Timoney}, \citenamefont {Weiss},
  \citenamefont {Gariglio}, \citenamefont {Warburton}, \citenamefont
  {Sch{\"o}nenberger}, \citenamefont {Renema}, \citenamefont {Zbinden},\ and\
  \citenamefont {Bussieres}}]{caloz2017}%
  \BibitemOpen
  \bibfield  {author} {\bibinfo {author} {\bibfnamefont {M.}~\bibnamefont
  {Caloz}}, \bibinfo {author} {\bibfnamefont {B.}~\bibnamefont {Korzh}},
  \bibinfo {author} {\bibfnamefont {N.}~\bibnamefont {Timoney}}, \bibinfo
  {author} {\bibfnamefont {M.}~\bibnamefont {Weiss}}, \bibinfo {author}
  {\bibfnamefont {S.}~\bibnamefont {Gariglio}}, \bibinfo {author}
  {\bibfnamefont {R.~J.}\ \bibnamefont {Warburton}}, \bibinfo {author}
  {\bibfnamefont {C.}~\bibnamefont {Sch{\"o}nenberger}}, \bibinfo {author}
  {\bibfnamefont {J.}~\bibnamefont {Renema}}, \bibinfo {author} {\bibfnamefont
  {H.}~\bibnamefont {Zbinden}}, \ and\ \bibinfo {author} {\bibfnamefont
  {F.}~\bibnamefont {Bussieres}},\ }\href {\doibase 10.1063/1.4977034}
  {\bibfield  {journal} {\bibinfo  {journal} {Appl. Phys. Lett.}\ }\textbf
  {\bibinfo {volume} {110}},\ \bibinfo {pages} {083106} (\bibinfo {year}
  {2017})}\BibitemShut {NoStop}%
\bibitem [{\citenamefont {Korzh}\ \emph {et~al.}(2014)\citenamefont {Korzh},
  \citenamefont {Walenta}, \citenamefont {Lunghi}, \citenamefont {Gisin},\ and\
  \citenamefont {Zbinden}}]{Korzh2014}%
  \BibitemOpen
  \bibfield  {author} {\bibinfo {author} {\bibfnamefont {B.}~\bibnamefont
  {Korzh}}, \bibinfo {author} {\bibfnamefont {N.}~\bibnamefont {Walenta}},
  \bibinfo {author} {\bibfnamefont {T.}~\bibnamefont {Lunghi}}, \bibinfo
  {author} {\bibfnamefont {N.}~\bibnamefont {Gisin}}, \ and\ \bibinfo {author}
  {\bibfnamefont {H.}~\bibnamefont {Zbinden}},\ }\href {\doibase
  10.1063/1.4866582} {\bibfield  {journal} {\bibinfo  {journal} {Appl. Phys.
  Lett.}\ }\textbf {\bibinfo {volume} {104}},\ \bibinfo {pages} {081108}
  (\bibinfo {year} {2014})}\BibitemShut {NoStop}%
\bibitem [{\citenamefont {Fung}\ \emph {et~al.}(2010)\citenamefont {Fung},
  \citenamefont {Ma},\ and\ \citenamefont {Chau}}]{fung2010}%
  \BibitemOpen
  \bibfield  {author} {\bibinfo {author} {\bibfnamefont {C.-H.~F.}\
  \bibnamefont {Fung}}, \bibinfo {author} {\bibfnamefont {X.}~\bibnamefont
  {Ma}}, \ and\ \bibinfo {author} {\bibfnamefont {H.}~\bibnamefont {Chau}},\
  }\href {\doibase 10.1103/PhysRevA.81.012318} {\bibfield  {journal} {\bibinfo
  {journal} {Physical Review A}\ }\textbf {\bibinfo {volume} {81}},\ \bibinfo
  {pages} {012318} (\bibinfo {year} {2010})}\BibitemShut {NoStop}%
\end{thebibliography}%
\end{document}